\def \tr {{\rm tr}\,}
\documentclass[twocolumn,showpacs,preprintnumbers,amsmath,amssymb]{revtex4}
\usepackage{epsf}


\begin{document}
\title{Gauge pathologies in singularity-avoidant spacetime foliations}
\author{C~Bona$^1$, T~Ledvinka$^2$, C~Palenzuela-Luque$^3$, J A~Pons$^4$ and
M.~\v Z\' a\v cek$^2$}

\affiliation{ $^1$ Departament de Fisica, Universitat de les Illes
Balears, Palma de Mallorca, Spain \\
$^2$ Institute of Theoretical Physics, Charles University, Prague,
Czech Republic \\
$^3$ Department of Physics and Astronomy,
Louisiana State University, Louisiana, USA \\
$^4$ Departament de F\'{\i}sica Aplicada, Universitat d'Alacant,
Alacant, Spain}

\begin{abstract}
The family of generalized-harmonic gauge conditions, which is
currently used in Numerical Relativity for its
singularity-avoidant behavior, is analyzed by looking for
pathologies of the corresponding spacetime foliation. The
appearance of genuine shocks, arising from the crossing of
characteristic lines, is completely discarded. Runaway solutions,
meaning that the lapse function can grow without bound at an
accelerated rate, are instead predicted. Black Hole simulations
are presented, showing spurious oscillations due to the well known
slice stretching phenomenon. These oscillations are made to
disappear by switching the numerical algorithm to a
high-resolution shock-capturing one, of the kind currently used in
Computational Fluid Dynamics. Even with these shock-capturing
algorithms, runaway solutions are seen to appear and the resulting
lapse blow-up is causing the simulations to crash. As a side
result, a new method is proposed for obtaining regular initial
data for Black Hole spacetimes, even inside the horizons.
\end{abstract}

\maketitle

\section{Introduction}

The choice of a coordinate system is mandatory when finding
solutions of Einstein's field equations. In the 3+1
approach~\cite{ADM}, this choice can be made in two steps:
\begin{itemize}
    \item The choice of a time coordinate. In geometrical terms,
    this amounts to select a specific foliation of spacetime by
    a family of spacelike hypersurfaces (the constant time
    slices).
    \item The choice of the space coordinates. In geometrical
    terms, this amounts to select a specific congruence of lines
    threading every slice of the time foliation (the time lines).
\end{itemize}
Most of the theoretical developments in the 3+1 approach are
space-invariant, that is independent of the choice of the space
coordinates. In this case, one can safely select as time lines
precisely the normal lines to the time slices (normal
coordinates), as we will do in what follows. In local adapted
coordinates, we will write then the line element as
\begin{equation}\label{line_element}
    ds^2 = -\alpha^2~dt^2 + \gamma_{ij}~dx^i~dx^j\,.
\end{equation}

The choice of a time coordinate, however, is not so simple. In
numerical simulations one must care about physical singularities
that could arise dynamically in gravitational collapse scenarios.
Care must be taken to avoid the time slices to hit the singularity
in a finite amount of coordinate time: otherwise the numerical
code will crash due to the singular terms.

This is why singularity avoidant slicing conditions are widely
used in Numerical Relativity. A first example is provided by the
maximal slicing condition,
\begin{equation}\label{maximal}
    \tr K = 0\,,
\end{equation}
where
\begin{equation}\label{Kij}
    K_{ij} = -\frac{1}{2\alpha}~\partial_t~\gamma_{ij}
\end{equation}
is the extrinsic curvature tensor of the time slices. Maximal
slicing has been successfully used in Black Hole simulations since
the early times~\cite{SY78}. The behavior of maximal slicing in
simple cases has been studied for more than thirty
years~\cite{Eetal73} and is still being studied
today~\cite{Reimann04}. The problem with maximal slicing is that
it leads to a second order elliptic equation for the lapse
function $\alpha$. This can be a complication in the current 3D
simulations, where the need of accuracy is consuming all the
available computational resources. But this is also a complication
from the theoretical point of view, because the resulting
evolution system (field equations plus coordinate conditions) is
then of a mixed elliptic-hyperbolic type.

A simpler alternative is provided by the generalized harmonic
condition~\cite{BM95}
\begin{equation}\label{harmgen}
    \partial_t~\alpha = - f \alpha^2~\tr K\,,
\end{equation}
where $f$ can be an arbitrary function of $\alpha$ ($f=1$ in the
original harmonic slicing~\cite{CR83, BM88}). It provides a first
order evolution equation for the lapse function which, when
combined with the field equations, leads to a hyperbolic evolution
system (weakly hyperbolic or strongly hyperbolic, depending on the
formulation) for non-negative choices of $f$.

The problem with condition (\ref{harmgen}) is that it can lead to
gauge-related pathologies. In Ref.~\cite{AMSST95}, this has been
seen to occur in Black Hole simulations for some particular cases.
In Ref.~\cite{Alcubierre97}, the generic case is considered;
starting from an analysis of the characteristic speeds, the
conclusion is that there can be two classes of shocks:
\begin{itemize}
    \item one which affects just the gauge degrees of freedom. A
    cure is suggested, consisting in the choice
    \begin{equation}\label{fAlcub}
    f = 1 + k/\alpha^2
    \end{equation}
    ($k$ being just an integration constant).
    \item another one which affects even the metric degrees of
    freedom (with $c$ as characteristic speed), for which there is no
    cure.
\end{itemize}
The claim concerning the existence of gauge shocks has been
recently strengthened~\cite{Alcubierre03} by providing a purely
kinematical derivation of the condition (\ref{fAlcub}),
independently of the field equations.

These are by no means rhetorical questions. We are aware that
Einstein's equations can be used for describing the propagation of
discontinuities arising either from the initial or the boundary
data. We are talking here about something stronger: shocks that
could arise dynamically even from smooth initial and boundary
data. This is a well known phenomenon in the Fluid Dynamics
domain: the non-linearity of the equations leads in some
circumstances to the crossing of the characteristic lines so that
a genuine shock (not just a contact discontinuity) appears. Can
this really happen in General Relativity?. Or, at least, can this
happen when using (\ref{harmgen}) as slicing condition?. We think
that this is a good opportunity to address these questions and
provide, as far as possible, definite answers.

\section{Are there metric shocks?}

\begin{figure}[t]
\begin{center}
\epsfxsize=8cm \epsfbox{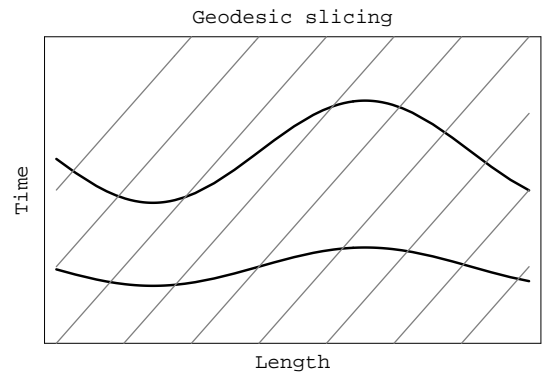}
\end{center}
\caption{The diagram shows a set of light rays as straight lines
with unit slope in a time-length diagram (geodesic slicing). Two
hypersurfaces (time slices) of an alternative slicing are also
shown. These hypersurfaces must be spacelike, so that the slope of
the tangent vectors is limited by that of the light rays. Notice
that the space distance among light rays, as measured on the time
slices, is no longer constant. But notice also that the order of
the sequence of light rays on each time slice is unaltered.}
\label{slicing}
\end{figure}

No, there are not metric shocks. To see why, let us start with the
space-invariant expression of the corresponding characteristic
speed (light speed)
\begin{equation}\label{light_speed}
    c = \frac{dl}{dt} = \sqrt{\gamma_{ij}~\frac{dx^i}{dt}~\frac{dx^j}{dt}}
    = \pm ~\alpha
\end{equation}
(length variation per unit coordinate time). We can always
consider a geodesic slicing, so that $\alpha=1$ and coordinate
time coincides with proper time. In this case, as depicted in
Fig.~\ref{slicing}, the characteristic lines (light rays) along
any given space direction can not even approach one another.

The question is whether these light rays can cross each other in
some alternative slicing. As shown in Fig.~\ref{slicing}, this can
never happen as long as the time slices are spacelike
hypersurfaces. Without characteristic crossing, there are not
genuine shocks (although contact discontinuities can appear from
initial or boundary data, as discussed in the Introduction).
Einstein's field equations can then be said to be, in the Applied
Mathematics jargon, linearly degenerate. The case of the gauge
degrees of freedom, which are not prescribed by the field
equations, will be discussed in the following sections.

But let us look back before at Ref.~\cite{Alcubierre97}, where a
coordinate-dependent expression for light speed was used as the
starting point. If we repeat the arguments of
Ref.~\cite{Alcubierre97}, but using instead the space-invariant
expression (\ref{light_speed}) as the starting point, we see that
the prediction of metric shocks disappears. The prediction of
gauge shocks remains, but condition (\ref{fAlcub}) for the
proposed cure is to be replaced by
\begin{equation}\label{fZ}
    f = k/\alpha^2\,,
\end{equation}
which no longer contains the original harmonic slicing ($f=1$) as
a particular case.

\section{Is there a kinematical reason for gauge shocks?}

No, there is no kinematical reason for gauge shocks. The argument
in Ref.~\cite{Alcubierre03} starts from the time-slicing equation
\begin{equation}\label{time_eq}
   \left[ g^{\mu\nu} + (1-\frac{1}{f})~
    \frac{d^\mu \phi~d^\nu \phi}{(-g^{\rho\sigma}d_\rho \phi~d_\sigma \phi)}
    \right]~\nabla_\mu d_\nu \phi = 0\,,
\end{equation}
which reduces to the generalized harmonic slicing condition
(\ref{harmgen}) in local adapted coordinates, where
\begin{equation}\label{adapted}
    \phi = t\,,\qquad \alpha = (-g^{\rho\sigma}d_\rho \phi~d_\sigma
    \phi)^{-1/2}\,.
\end{equation}
The argument proceeds then by transforming the quasilinear
second-order equation (\ref{time_eq}) into a first order system
and studying the characteristic speeds, without recourse to the
field equations~\cite{Alcubierre03}.

We will take here a more direct approach, looking for the
characteristic hypersurfaces of the original equation
(\ref{time_eq}). To simplify the analysis, we will adapt the
general coordinates in (\ref{time_eq}) to the (generic) initial
data. Starting from a given spacelike slice $\Sigma$, we will
choose then as initial conditions
\begin{equation}\label{initial_phi}
    \phi\mid_\Sigma = t_0\,,\qquad \partial_k\,\phi\mid_\Sigma = 0\,,
    \qquad \partial_t\,\phi\mid_\Sigma = \alpha\,.
\end{equation}
Notice that this is by no means a restriction on the initial data:
it is rather the result of adapting the arbitrary spacetime
coordinates to any specific choice of initial data. Moreover, this
kind of adapted coordinates is precisely the one that is currently
used in numerical simulations, where the equation (\ref{time_eq})
is always used in the form (\ref{harmgen}).

It is trivial now to look for the characteristic hypersurfaces.
All second partial derivatives on $\Sigma$ can be obtained from
(\ref{initial_phi}) with the only exception of the second time
derivative, which can be obtained from (\ref{time_eq}) unless
\begin{equation}\label{charact}
    \frac{1}{f} ~g^{00}\mid_\Sigma = 0\,,
\end{equation}
which would then imply that $\Sigma$ is a characteristic
hypersurface. For any finite choice of $f$, condition
(\ref{charact}), which can be written in an invariant way as
\begin{equation}\label{cones}
    g^{\mu\nu}\partial_\mu \phi\,\partial_\nu \phi\mid_\Sigma = 0\,,
\end{equation}
amounts to the statement that $\Sigma$ is a null hypersurface.
This means that the characteristic speed coincides again with
light speed. It follows that the discussion of the preceding
section applies here again to show that there can not be any
crossing of characteristics and hence no genuine shocks can
develop.

To put it in a different way: equation (\ref{time_eq}) can always
be used for constructing a new foliation in a given (regular)
spacetime provided that the resulting slices remain spacelike.
Then, the time slicing itself will be well defined, although the
behavior of the gauge-related dynamical fields may depend on the
field equations, as we will see below.

\section{Are there dynamical gauge shocks?}

There are certainly gauge pathologies of a dynamical origin, but
not genuine shocks. In order to analyze this point, it is
convenient to consider the time derivative of the generalized
harmonic slicing condition (\ref{harmgen}). This requires using
the field equations to provide the time evolution of $\tr K$ (the
arguments in this section will be then of a dynamical nature).

We will choose for simplicity the BSSN formulation~\cite{SN95,
BS99}, although all the current formalisms share the same physical
solutions. In the vacuum case, we get after some algebra
\begin{equation}
  (1/\tilde{f})~\partial_{tt}~\alpha - \triangle~\alpha =
 -~[~\alpha ~\tr(K^2)-\tilde{f}'~(\tr K)^2~]\,,
\label{alpha_master}
\end{equation}
where we have noted for short
\begin{equation}\label{ftilde}
    \tilde{f} \equiv f~\alpha^2\,,\qquad \tilde{f}'
    \equiv \frac{\partial\,\tilde{f}}{\partial\,\alpha}\,.
\end{equation}

Equation (\ref{alpha_master}) can be interpreted as a generalized
wave equation for the lapse function. The left-hand-side terms
correspond to the principal part, and it is clear that the
characteristic speed is given by
\begin{equation}\label{gauge_speed}
    v_G = \pm \sqrt{f}\,\alpha
\end{equation}
(gauge speed). We derived condition (\ref{fZ}) by using precisely
(\ref{gauge_speed}) as the starting point for the argument
presented in Ref.~\cite{Alcubierre97}: it amounts to the
requirement of a constant gauge speed, that is
\begin{equation}\label{vg_constant}
    v_G = {\rm constant}~\Leftrightarrow~\tilde{f}' = 0\,.
\end{equation}

The right-hand-side terms in (\ref{alpha_master}) can be
interpreted as non-linear source terms for the evolution of the
lapse function. Allowing for (\ref{harmgen}), $\tr K$ is
proportional to the time derivative of $\alpha$, so that there is
a non-linear coupling that can lead to runaway (growing without
bound, at an increasing rate) solutions.

An extreme example of this gauge pathology, in which transport
plays not role at all, is provided by space-homogeneous and
isotropic solutions, like the ones used in Cosmology, where all
space derivatives vanish and the extrinsic curvature is
proportional to the metric. One has then
\begin{equation}\label{runaway}
\partial_{tt}~\alpha =
 \tilde{f}\,(\tilde{f}'-\frac{\alpha}{3})\,(\tr K)^2 =
 \frac{1}{\tilde{f}}\,(\tilde{f}'-\frac{\alpha}{3})\,(\partial_{t}\,\alpha)^2\,,
\end{equation}
so that $\alpha$ can grow without bound, with a driving force
proportional to the square of the growing rate, unless
\begin{equation}\label{fZZ}
    \tilde{f}' \leq \frac{\alpha}{3}\qquad\Leftrightarrow\qquad
    f = n + \frac{k}{\alpha^2}\qquad (n\leq \frac{1}{6})\,.
\end{equation}

Coming back to the general case, it is clear that condition
(\ref{vg_constant}) could be interpreted in two alternative ways
\begin{itemize}
    \item Ensuring a constant gauge speed, so that no
    characteristic crossing can occur.
    \item Ensuring the negative sign of the source terms in
    (\ref{alpha_master}), so that no runaway solutions can occur.
\end{itemize}
This ambiguity follows from the very nature of the argument
proposed in Ref.~\cite{Alcubierre97}, where the main idea is to
include the source terms in the discussion about the behavior of
characteristic fields. The $\tilde{f}$-dependent term in the
right-hand-side of (\ref{alpha_master}) would then be considered
as a non-linear contribution to the transport (left-hand-side)
terms.

But we do not see much difference between this particular
quadratic term and the other ones in the right-hand-side of
(\ref{alpha_master}), independently of their origin. The example
we presented before shows instead how these other source terms
actually contribute to the regularity condition (\ref{fZZ}) by
providing an upper bound which was missing in (\ref{vg_constant}).

Our conclusion is that any pathological behavior of the
gauge-dependent degrees of freedom, which manifests itself as an
unbounded growth of either the lapse function or its first
derivatives, can be interpreted consistently as the effect of the
non-linear source terms in the evolution equations. We will test
this conclusion against simple numerical Black-Hole simulations in
what follows.

\section{Gauge pathologies in numerical Black-Hole simulations}

We will focus here on three-dimensional numerical simulations of a
spherically symmetric Black Hole. As far as we are interested in
the strong field region, the Black Hole interior is not excised
from the numerical grid. Regular initial data are obtained by
using a novel 'Free Black Hole' approach, which relies on the well
known 'no hair' theorems (see Appendix I for details). This new
approach is also being applied independently in the context of the
BSSN formalism~\cite{Denis}.

We will use in our simulations the first order version of the Z4
evolution formalism~\cite{Z4,Z48} in normal coordinates (zero
shift). Apart from standard centered finite-difference algorithms,
we will use at some point the MMC method. This is a
high-resolution shock-capturing (HRSC) method which combines the
monotonic-centered (MC) flux limiter~\cite{vanLeer77} with
Marquina's flux formula~\cite{Marquina} (see Appendix II for
details). The use of these advanced HRSC methods  is also new in
3D Black Hole simulations, although it is crucial for our
arguments about shocks.

\subsection{Slice stretching}

\begin{figure} [b]
\caption{Lapse collapse for the '1+log' slicing : the lapse values
are shown every $0.5M$. The lapse gets very close to zero in the
innermost region. A standard finite difference algorithm is used
in a cubic cartesian grid (the space direction in the plots
corresponds to the grand diagonal of the cube). Spurious
oscillations appear at about $t=6M$ in the zones where the slope
is changing more abruptly.}
\begin{center}
\epsfxsize=8.5cm \epsfbox{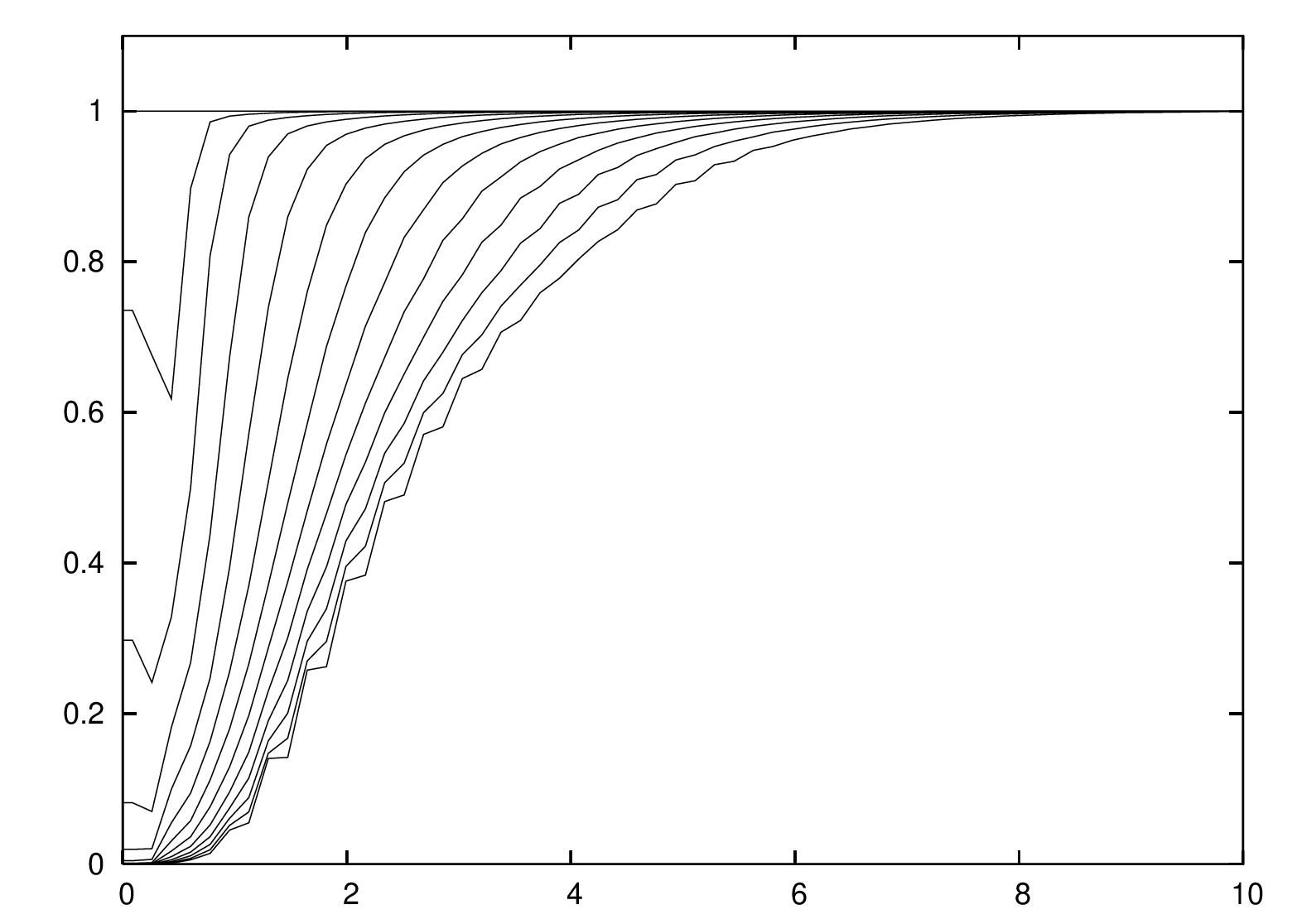}
\end{center}
\label{Gibbs}
\end{figure}

The first stages of the simulation show a collapse of the space
volume element which, allowing for the singularity avoidance
properties of the gauge conditions (\ref{harmgen}), translates
itself into a collapse of the lapse function $\alpha$. This lapse
collapse can be slower or faster, depending on our choice of $f$.
We will present here our results for the '1+log' case
\begin{equation}\label{1+log}
    f = \frac{2}{\alpha}\,,
\end{equation}
although no qualitative difference is detected for similar
choices, like $f=2$ for instance.

\begin{figure} [t]
\begin{center}
\epsfxsize=8cm \epsfbox{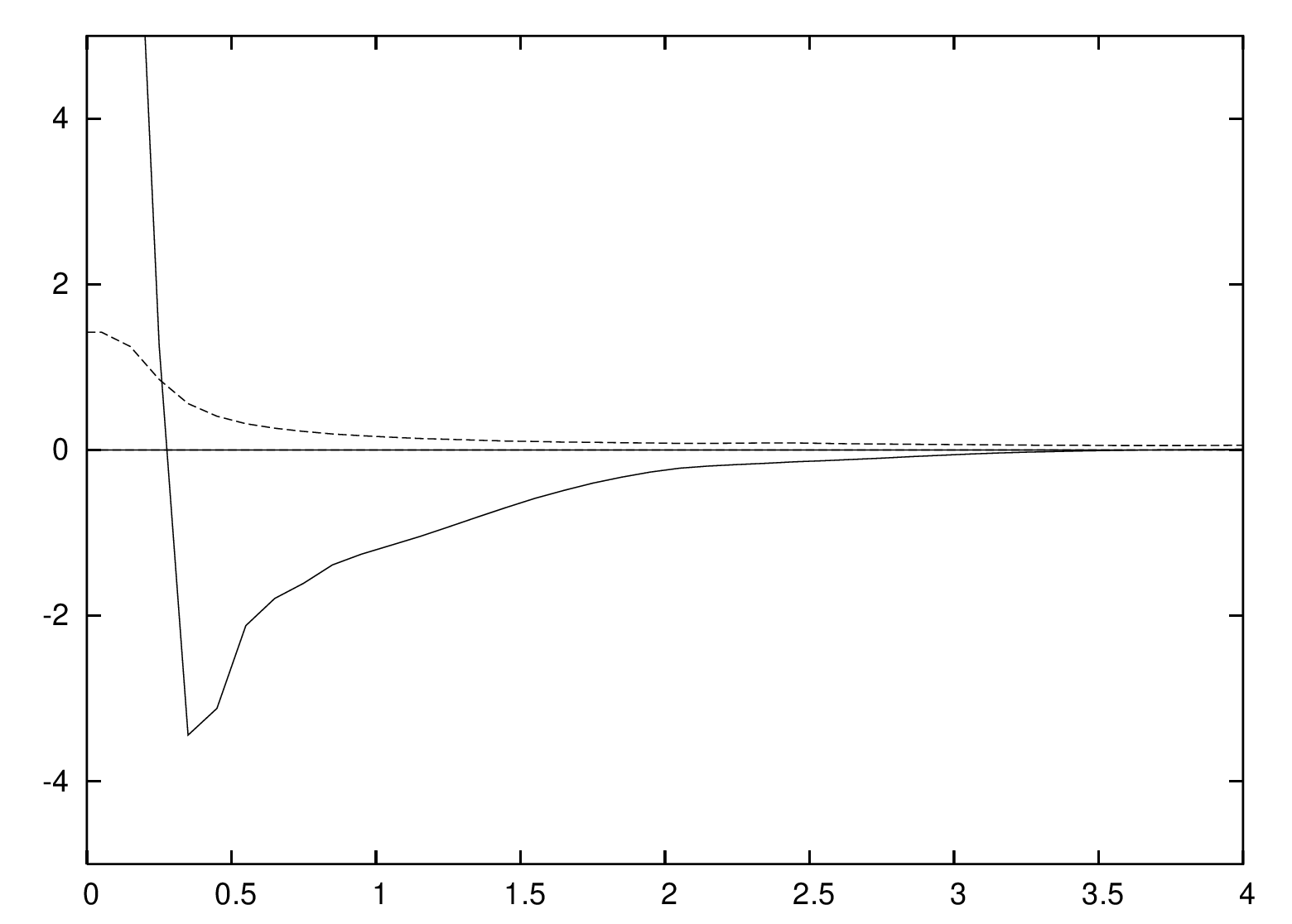}
\end{center}
\caption{Plots of $K_{xx}$ (solid line) and $\tr K$ (dotted line)
along the $x$ axis, corresponding to the 1+log slicing at $t=6M$.
Negative values of $K_{xx}$ correspond to the increasing of
$\gamma_{xx}$, that is a radial stretching of the time slice.
Notice that this is compatible with an overall collapse pattern,
as shown by the positivity of $\tr K$: the radial stretching is
then compensated by the collapse along the angular directions. The
profile of $\tr K$ is smooth, without any sign of shocks.}
\label{stretching}
\end{figure}

\begin{figure} [b]
\caption{Same as Figure~\ref{Gibbs}, but using the MMC numerical
algorithm for the space discretization, as described in Appendix
II. The spurious oscillations have completely disappeared. Slice
stretching is still there (Figure~\ref{stretching} corresponds
actually to the last plot), but it is no longer a problem for
numerical simulations.}\label{smoothed}
\begin{center}
\epsfxsize=8.5cm \epsfbox{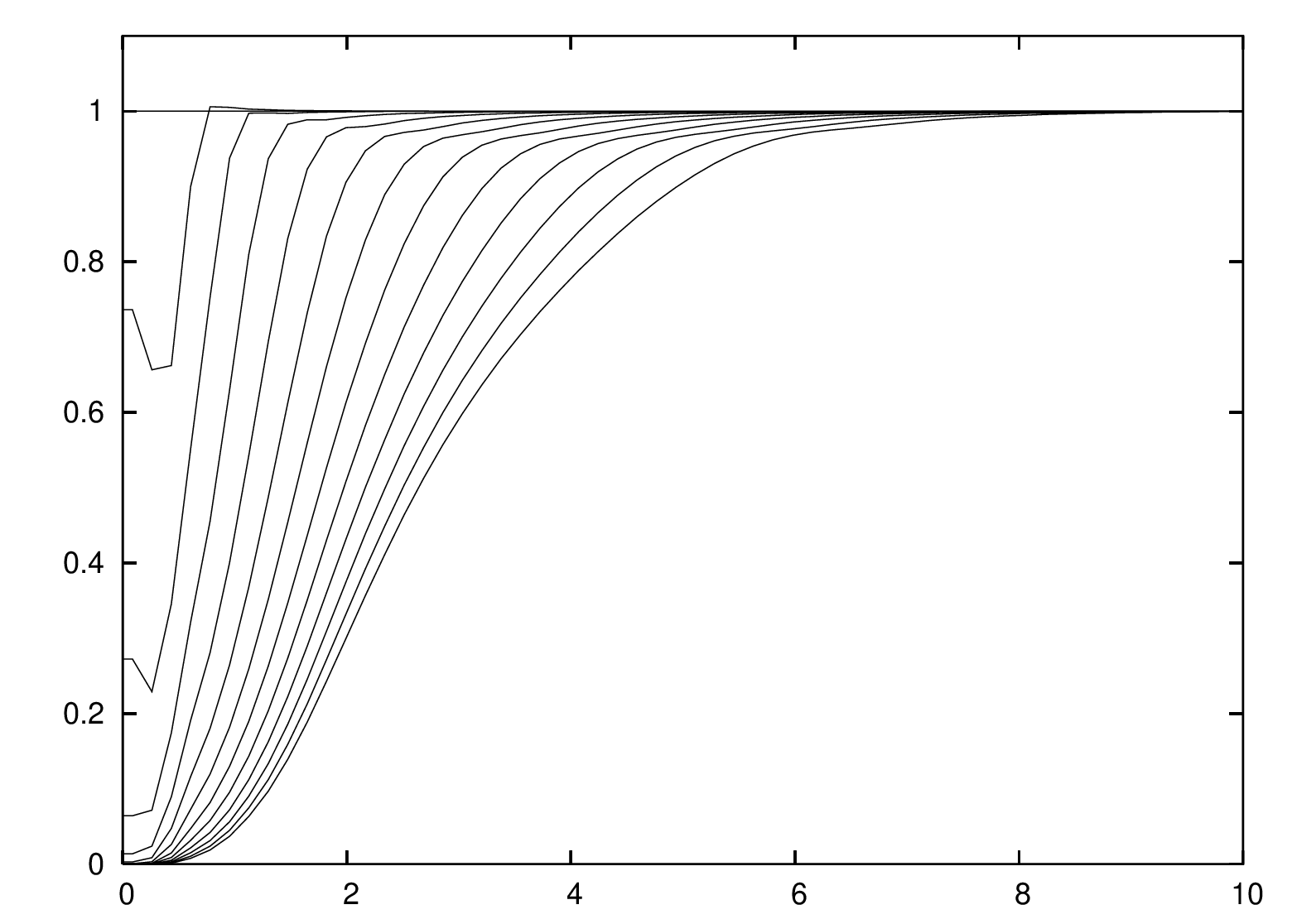}
\end{center}
\end{figure}

Let us see what happens after some time. We see in
Fig.~\ref{Gibbs} how high-frequency oscillations start appearing
in the regions where the lapse slope is changing more abruptly.
This is a numerical effect: it is the analogous of the well-known
Gibbs phenomenon, which appears when the tool one is using (the
standard finite difference method in our case) is unable to
resolve the higher frequency modes of some dynamical field.

In order to see where those annoying high-frequency modes came
from, let us take a look at behavior of the extrinsic curvature,
as displayed in Fig.~\ref{stretching}. What we see there is the
well known 'slice stretching' phenomenon: the Black Hole is
sucking both the numerical grid~\cite{ST86,ADMSS95} and the space
slice~\cite{Reimann04,RB04}, producing as a result steep profiles
along the radial direction. High-frequency modes are the expected
to appear in the regions with abrupt slope changes in some
dynamical field.

The same kind of spurious oscillations currently appears in
Computational Fluid Dynamics simulations, in regions where genuine
shocks are developing. The cure is to use advanced HRSC numerical
methods that can deal with shocks without altering the
monotonicity of the dynamical fields. We will use in what follows
the MMC numerical algorithm, as described in Appendix II, which is
one of such HRSC methods. We are aware that we are not dealing
here with real shocks (the $\tr K$ profile in
Fig.~\ref{stretching} is clearly smooth). We are instead dealing
with, let us say, 'numerical shocks' as far as our numerical grid
is just unable to resolve the higher frequency modes associated
with slice stretching.

The effect of switching to the MMC numerical method is definitive,
as it can be seen on Fig.~\ref{smoothed}. No trace of the spurious
oscillations remains. There is nothing either in the lapse or in
the $\tr K$ profiles that could be interpreted as a shock. The
conclusion is obvious: slice stretching does not produce any real
shock, it may produce just 'numerical shocks' that disappear when
improving the discretization algorithm.

\subsection{Runaway solutions}

\begin{figure} [t]
\begin{center}
\epsfxsize=8.5cm \epsfbox{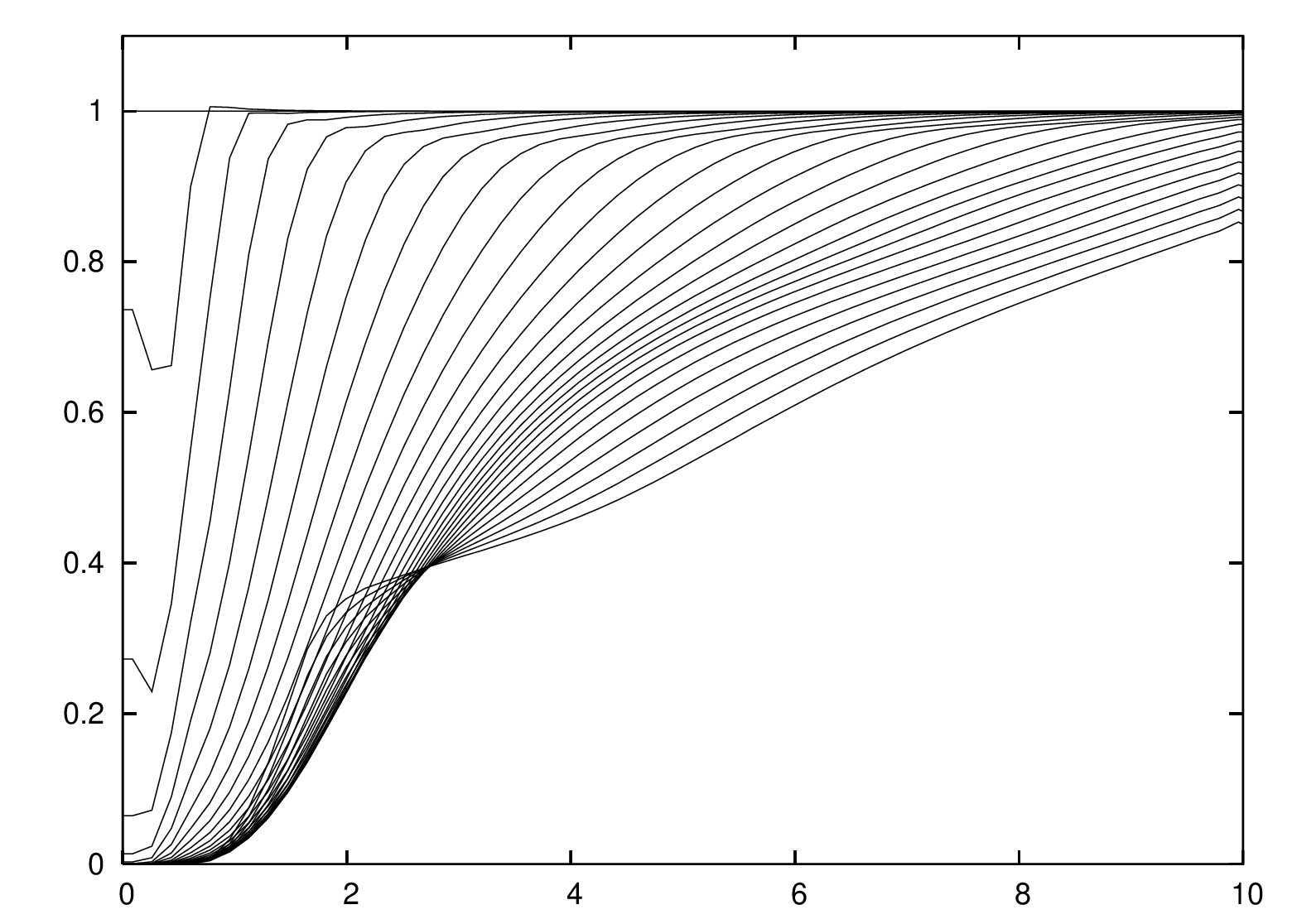}
\end{center}
\caption{Same as Figure~\ref{smoothed}, but with the simulations
running up to $t=14M$. No spurious oscillations appear anymore.
However, the lapse collapse stops and a sort of rebound starts
near the frozen innermost region. The lapse looks just going
backwards there. The simulation is going to crash by a lapse blow
up.}\label{rebounding}
\end{figure}

Let us see now what happens when allowing our simulations to
proceed for a longer time (about $t=14M$ in our case, but the
value will depend on the specific gauge choice). We see in
Fig.~\ref{rebounding} that the spurious oscillations do not show
up anymore. What we see instead is a sort of rebound of the lapse,
which is the prelude of a blow up that will crash the numerical
simulation. We have stopped the simulation in
Fig.~\ref{rebounding} before the crash occurs, so that two
different interpretations can be tentatively considered
\begin{itemize}
    \item A gauge shock is forming and starts propagating
    backwards in the region near to $\rho=2M$.
    \item The growing of the lapse is triggering a runaway
    solution that will result in a lapse blow up.
\end{itemize}

\begin{figure} [t]
\begin{center}
\epsfxsize=8.5cm \epsfbox{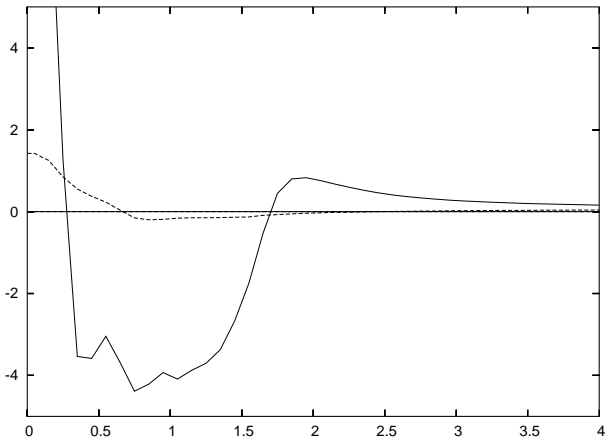}
\end{center}
\caption{Same as Fig.~\ref{stretching}, but now for $t=14M$,
corresponding to the last plot in Fig.~\ref{rebounding}. The
dynamics in the innermost region is frozen (the lapse is fully
collapsed there). In the region around $\rho=2M$, however,
positive values of $K_{xx}$ appear, corresponding to a radial
collapse, whereas $\tr K$ becomes negative, corresponding to a
global expansion pattern. This is just the behavior opposite to
slice stretching, and it explains the lapse rebound that can be
seen in Fig.~\ref{rebounding}. Notice that the collapse pattern
(positive $trK$ values) is recovered in the outermost region,
where the lapse collapse proceeds as expected.}\label{rebound}
\end{figure}

In order to discriminate between these two alternatives, let us
take a look to Fig.~\ref{rebound}, where we have plotted the same
extrinsic curvature components as in Fig.~\ref{stretching}. As a
word of caution, let us remember that the lapse in the innermost
region is yet collapsed, so that the dynamics is frozen there.
This means that the features we see in the innermost region in
Fig.~\ref{rebound} correspond to an earlier stage (measured in
proper time) than what we see around $\rho=2M$, which is the
region we are going to analyze now.

What we see there is just the opposite that what we saw in
Fig.~\ref{stretching}: $\tr K$ is now negative, so that we have an
overall expansion pattern. But the radial component ($K_{xx}$
along the $x$ direction) is collapsing. This explains why the
whole picture in Fig.~\ref{rebounding} looks like moving backwards
in time in the rebounding region.

Notice that the $\tr K$ profile in Fig.~\ref{rebound} is clearly
smooth and well-resolved in the region around $\rho=2M$, so that
the shock interpretation can be excluded. Moreover, as far as we
are using a HRSC numerical method, no blow up would be expected
from the presence of shocks. In Computational Fluid Dynamics these
methods are currently used for dealing with genuine shocks without
crashing because of that. The blow up must be actually caused by
something much stronger than a mere shock: a runaway solution that
is triggered by the negative values of $\tr K$, which correspond
to an overall expansion behavior.

The appearance of runaway solutions is to be expected when
selecting the '1+log' (\ref{1+log}) or any similar choice among
the singularity-avoidant family of slicing conditions
(\ref{harmgen}). If we decompose the extrinsic curvature tensor
$K_{ij}$ into its irreducible components
\begin{equation}\label{K_decomp}
    A_{ij} \equiv K_{ij}- \frac{\tr K}{3}~\gamma_{ij}\,,\qquad
    K \equiv \tr K
\end{equation}
(shear tensor and expansion factor, respectively), we see that the
source terms in the lapse equation (\ref{alpha_master}) can be
expressed as
\begin{equation}\label{driving}
    - \alpha~(\tr A^2) + (2-\frac{\alpha}{3})~K^2\,,
\end{equation}
which can easily get a positive sign in collapsing scenarios
($\alpha$ small) unless the expansion factor $K$ is kept to low
values when compared with the shear tensor.

We conclude that runaway solutions, not gauge shocks, can be a
real problem in simulations that, like in the Black Hole cases,
make use of the singularity-avoidant slicing conditions
(\ref{harmgen}). To find a convenient cure for this gauge
pathology (excising the interior region, using a dynamical shift,
selecting other values of $f$, and so on) goes beyond the purpose
of this paper.

\renewcommand{\theequation}{I.\arabic{equation}}
\setcounter{equation}{0}
\section*{Appendix I: Free Black Hole initial data}

Based in the idea that the interior region of a Black Hole has no
causal physical influence on the exterior one (no-hair theorems),
we can devise a simple way of obtaining regular initial data for
Black Hole spacetimes:
\begin{itemize}
    \item Solve the energy and momentum constraints for the initial
    data. This currently leads to a singular space metric.
    \item Replace the metric in the interior regions by any smooth
    extension of the exterior geometry. Energy and momentum
    constraints will be no longer fulfilled inside the horizons
    ('Free Black Hole' data).
\end{itemize}

In the simplest case, we can consider time-symmetric vacuum
initial data with a conformally flat space metric, that
is
\begin{equation}\label{TSCF}
    K_{ij}\mid_{t=0} ~=~ 0\,,\qquad
    \gamma_{ij}\mid_{t=0} ~=~ \Psi^4~\delta_{ij}\,,
\end{equation}
so that the momentum constraint is trivially satisfied and the
energy constraint amounts to require that $\Psi$ be an harmonic
function of the Euclidean metric.

\begin{figure} [t]
\begin{center}
\epsfxsize=8cm \epsfbox{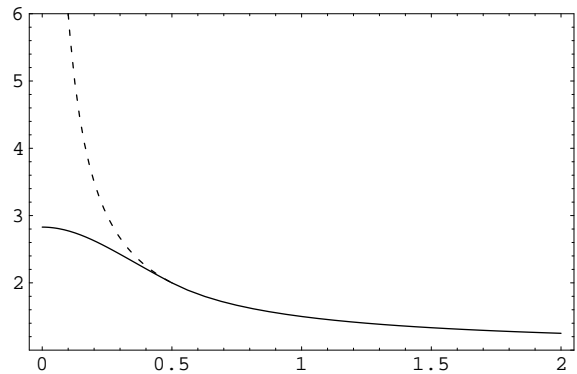}
\end{center}
\caption{Plot of a conformal factor $\Psi$ providing
time-symmetric conformally-flat initial data for a free
Schwarzschild Black Hole. Values on the $x$ axis correspond to the
isotropic radial coordinate $\rho$, measured in units of $M$. The
singular solution in the interior region (dashed line) is replaced
by a regular one (continuous line). Both expressions coincide in
the exterior region. Notice that the matching is smooth at the
apparent horizon ($\rho = M/2$). }\label{matching}
\end{figure}

Free initial data can be obtained then for a spherically symmetric
Black Hole with mass $M$ by starting from the singular conformal
factor
\begin{equation}\label{Psi_pot}
    \Psi = 1+\frac{M}{2\rho}\,,
\end{equation}
and replacing its value inside the horizon (dotted line in
Fig.~\ref{matching}) by any smooth function (continuous line in
Fig.~\ref{matching}).

In the Z4 formalism, energy and momentum constraint violations
will cause the growing of the supplementary fields $\Theta$ and
$Z_k$, which non-zero values correspond to a departure of true
Einstein's solutions. But these fields, which are the 3+1
components of the four-vector $Z_{\mu}$ are known to
verify~\cite{Z4}
\begin{equation}\label{boxeq}
    \Box Z_{\mu} + R_{\mu\nu} Z^{\nu} = 0\,,
\end{equation}
so that the spurious non-zero values propagate with light speed.
The exterior geometry, then, can not be affected by constraint
violations inside the horizon (at least not at the continuum
level).

\renewcommand{\theequation}{II.\arabic{equation}}
\setcounter{equation}{0}
\section*{Appendix II: The MMC high-resolution method}

High-Resolution Shock-Capturing (HRSC) numerical methods can be
applied to strongly-hyperbolic first-order systems of balance
laws. The balance-law structure means that the evolution equations
for the array $\mathbf{u}$ of dynamical fields can be written as
\begin{equation}\label{balance_law}
    \partial_t ~\mathbf{u} + \partial_k ~\mathbf{F}^k
    = \mathbf{S} \,,
\end{equation}
where both the Flux $\mathbf{F}^k$ and the Source $\mathbf{S}$
terms depend algebraically on the fields, but not on their
derivatives:
\begin{equation}\label{fluxes&Sources}
    \textbf{F}^k = \textbf{F}^k(\textbf{u})\,,
    \qquad \textbf{S} = \textbf{S}(\textbf{u})\,.
\end{equation}
The system (\ref{balance_law}) will be strongly hyperbolic if and
only if for every space direction $\vec{n}$, the corresponding
characteristic matrix
\begin{equation}\label{jacobi}
    n_k\mathbf{A}^k = n_k\frac{\partial \mathbf{F}^k}{\partial
    \mathbf{u}}
\end{equation}
has real eigenvalues $\mathbf{v}$ (characteristic speeds) and a
complete set $\mathbf{w}$ of eigenvectors~\cite{KL89}.

\begin{figure} [b]
\caption{\label{reconstruction} Starting from the values of the
fluxes at the grid nodes, a linear reconstruction is done inside
every elementary cell. Numerical discontinuities appear at every
interface (dotted lines) between the left and right values (arrows
and dots, respectively). Notice that the original function was
monotonically decreasing: all the slopes are negative. In the
proposed reconstruction, however, both the left interface sequence
(at $i-3/2$) and the right interface one (at $i+3/2$) show local
extreme values that break the monotonicity of the original
function.}
\begin{center}
\epsfxsize=8 cm \epsfbox{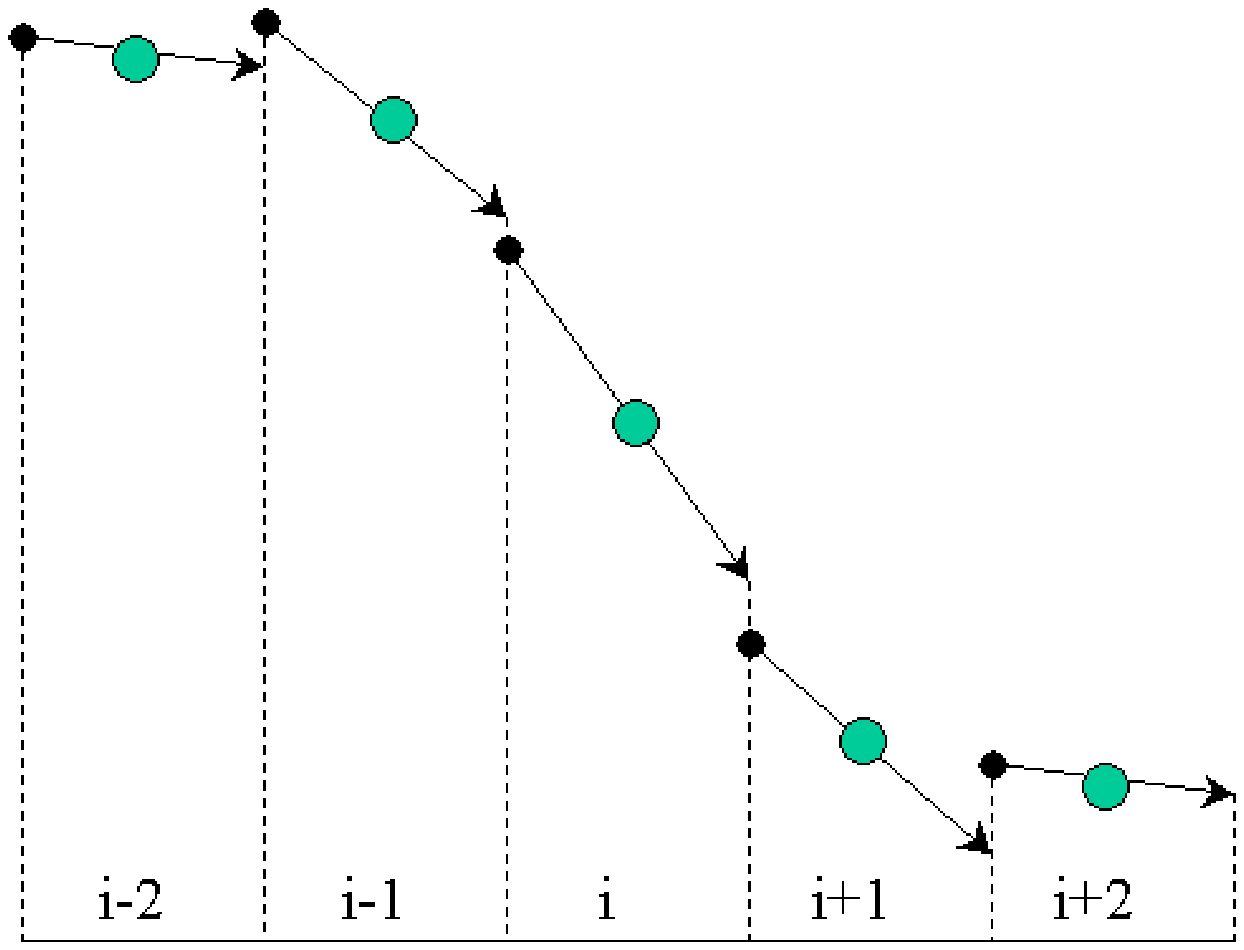}
\end{center}
\end{figure}

The balance law form (\ref{balance_law}) is specially suited for
the method-of-lines (MoL) discretization~\cite{MoL}. In this
method, there is a clear-cut separation between space and time
discretization. As a consequence, the source terms contribute in a
trivial way to the space discretization. The non-trivial
contribution comes instead from the space derivatives of the Flux
terms, which are usually discretized as follows
\begin{eqnarray}\label{balance_FD}\nonumber
    \partial_k ~\mathbf{F}^k &\sim&
    \frac{1}{\Delta x}~[\textbf{F}^x_{i+1/2}-\textbf{F}^x_{i-1/2}]
    + \frac{1}{\Delta y}~[\textbf{F}^y_{j+1/2}-\textbf{F}^y_{j-1/2}]\\
    &+& \frac{1}{\Delta z}~ [\textbf{F}^z_{k+1/2}-\textbf{F}^z_{k-1/2}]
    \,.
\end{eqnarray}
The half-integer indices correspond to the grid '\,interfaces',
which are supposed to be placed halfway between neighboring grid
nodes.

Every different way of computing the interface Fluxes in terms of
the values of the fields at the grid nodes will lead to a specific
numerical algorithm. A common feature of all these algorithms is
that the information of every grid node is used for providing
one-sided predictions at the neighbor interfaces in a consistent
way. For instance, one can define
\begin{equation}\label{fRL}
  F^R_{i-1/2} = F_i - \frac{\Delta x}{2} \Delta_i\,, \qquad
  F^L_{i+1/2} = F_i + \frac{\Delta x}{2} \Delta_i\,.
\end{equation}
Any numerical algorithm must then provide two basic elements:
\begin{itemize}
    \item A prescription for computing the slopes
    $\Delta_i$ which must be used in the
    'reconstruction' process (\ref{fRL}).
    \item A 'Flux formula' that provides a unique value of the
    interface Flux $F_{i+1/2}$ from the one-sided predictions
    $\{F^R_{i+1/2},F^L_{i+1/2}\}$.
\end{itemize}

\begin{figure} [b]
\caption{Same as Figure~\ref{reconstruction}, but using the
monotonic-centered reconstruction. Notice that the interface
values are bounded now between the neighbor nodes, so that
monotonicity is preserved both for the left values (arrows) and
for the right ones (dots) at every interface (dotted lines).}
\label{mcrecon}
\begin{center}
\epsfxsize=8 cm \epsfbox{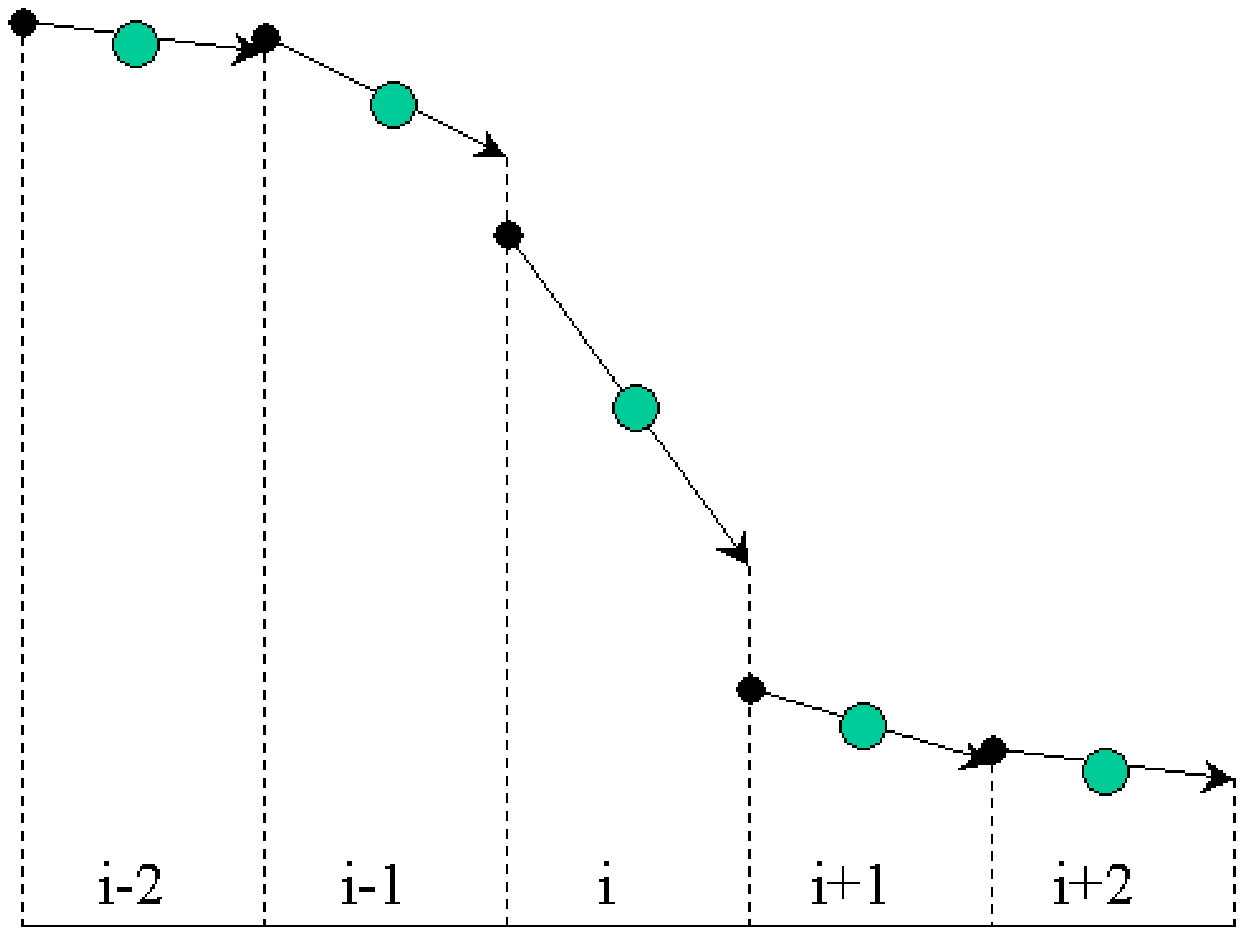}
\end{center}
\end{figure}

We can see in Fig.~\ref{reconstruction} an example of the
reconstruction process for the centered choice
\begin{equation}\label{slopec}
    \Delta^C_i = \frac{1}{2}~(F_{i+1}-F_{i-1})\,.
\end{equation}
The outcome consists in two sequences of one-sided predictions at
every interface (dots and arrows). Notice that the monotonicity of
the original Flux sequence is not preserved by the one-sided
predictions in the regions with abrupt slope changes. This can
lead to spurious oscillations in the numerical solution, even for
smooth dynamical fields, like the one displayed in
Fig.~\ref{reconstruction}. These spurious oscillations are visible
in the Black Hole simulation shown in Fig.~\ref{Gibbs}.

The MMC method uses instead a monotonicity-preserving
prescription, which can be obtained by using the 'monotonic
centered' (MC) slope~\cite{vanLeer77}
\begin{equation}\label{slopemc}
    \Delta^{MC}_i \equiv
    minmod\{\,2(F_i-F_{i-1}),\,2(F_{i+1}-F_{i}),\,\Delta^{C}_i\}
\end{equation}
instead of (\ref{slopec}). The $minmod$ function is defined as
usual:
\begin{itemize}
    \item If all the arguments have the same sign, then it selects
    the one with smaller absolute value.
    \item If one of the arguments has different sign than the others,
    then it is zero.
\end{itemize}
In this way, the slopes are limited in order to avoid spurious
oscillations. The rule is that interface values must lie between
their neighbor node values (see Figure~\ref{mcrecon}).

The second ingredient in the MMC method is the use of Marquina's
Flux formula~\cite{Marquina}. The prescription to calculate the
interface values $\mathbf{F}_{i+1/2}$ can be stated as follows (we
have simplified the original formula by adapting it to the
quasilinear evolution systems one gets from Einstein's field
equations):
\begin{itemize}
    \item Decompose both one-sided predictions
    $(\mathbf{F}^L\,,~\mathbf{F}^R)$ as linear combinations of
    the set of characteristic fields $\mathbf{w}$. Note that the
    coefficients in these combinations are not necessarily constant:
    we must use in general a different set of values on
    each side of the interface.
    \item Project the forward prediction $\mathbf{F}^L$, by
    suppressing any components $\mathbf{w}$ corresponding to negative
    characteristic speeds (forward projection $\mathbf{F}_{F}$).
    \item Project the backward prediction $\mathbf{F}^R$, by
    suppressing any components $\mathbf{w}$ corresponding to positive
    characteristic speeds (backward projection $\mathbf{F}_{B}$).
    \item Add these upwind-projected values at the given interface,
    that is
    \begin{equation}\label{upwindw}
    \mathbf{F}_{i+1/2} ~=~ \mathbf{F}_{F} + \mathbf{F}_{B}\,.
    \end{equation}
\end{itemize}
One is taking in this way the positive speed components from the
previous cell and the negative speed components from the next one.
Notice that we must assume for consistency a definite sign for the
characteristic speed (just the sign, not even the value). If the
sign changes between both sides of the given interface (when using
a super-luminal shift, for instance), then another combination
must be taken instead of the simple upwind projections presented
here (see Ref.~\cite{Marquina} for the details).

{\em Acknowledgements: This work has been supported by the EU
Programme 'Improving the Human Research Potential and the
Socio-Economic Knowledge Base' (Research Training Network Contract
HPRN-CT-2000-00137), by the Spanish Ministerio de Ciencia y
Tecnologia through the research grant number BFM2001-0988 and by a
grant from the Conselleria d'Innovacio i Energia of the Govern de
les Illes Balears.}

\bibliographystyle{prsty}

\begin{thebibliography}{99}

\bibitem{ADM} R.~Arnowit, S.~Deser and C.~W.~Misner.
        In: \textit{Gravitation:
        an introduction to current research}, ed. by L.~Witten,
        Wiley, (New York 1962).

\bibitem{SY78} L.~Smarr and J.~W.~York,
        Phys.~Rev. {\bf D17}, 1945 (1978),
        Phys.~Rev. {\bf D17}, 2529 (1978).

\bibitem{Eetal73} F.~Estabrook et al.,
        Phys.~Rev. {\bf D7}, 2814 (1973).

\bibitem{Reimann04} B.~Reimann,
        gr-qc/0404118, (2004).

\bibitem{BM95} C.~Bona, J.~Mass\'o, E.~Seidel
        and J.~Stela, Phys.~Rev.~Lett. {\bf 75} 600 (1995).

\bibitem{CR83} Y.~Choquet-Bruhat and T.~Ruggeri,
         Comm.~Math.~Phys. {\bf 89}, 269 (1983).

\bibitem{BM88} C.~Bona and J.~Mass\'o,
         Phys.~Rev. {\bf D38}, 2419 (1988).

\bibitem{AMSST95} P.~Anninos et al,
        Phys.~Rev. {\bf D52} 2059 (1995).

\bibitem{Alcubierre97} M.~Alcubierre,
         Phys.~Rev. {\bf D55}, 5981 (1997).

\bibitem{Alcubierre03} M.~Alcubierre,
         Class.~Quantum Grav. {\bf 20}, 607 (2003).

\bibitem{SN95} M.~Shibata and T.~Nakamura,
        Phys.~Rev. {\bf D52} 5428 (1995).

\bibitem{BS99} T.~W.~Baumgarte and S.~L.~Shapiro,
        Phys.~Rev. {\bf D59} 024007 (1999).

\bibitem{Denis} D.~Pollney, private communication.

\bibitem{Z4} C.~Bona, T.~Ledvinka, C.~Palenzuela, M.~\v Z\'a\v cek,
        Phys.~Rev. {\bf D67} 104005 (2003).

\bibitem{Z48} C.~Bona, T.~Ledvinka, C.~Palenzuela, M.~\v Z\'a\v cek,
         Phys.~Rev. {\bf D69} 064036 (2004).

\bibitem{vanLeer77} B.~van~Leer,
        J.~Comput.~Phys. {\bf 23}, 276 (1977).

\bibitem{Marquina} R.~Donat and A.~Marquina,
        J.~Comp.~Phys. {\bf 125}, 42 (1996).

\bibitem{ST86} S.~L.~Shapiro and S.~A.~Teukolsky,
        In: \textit{Dynamical Spacetimes and Numerical Relativity},
        ed. by J.~M.~Centrella,
        Cambridge University Press, (Cambridge, UK 1986).

\bibitem{ADMSS95} P.~Anninos et al,
        Phys.~Rev. {\bf D51} 5562 (1995).

\bibitem{RB04} B.~Reimann and B.~Br\"{u}gmann,
        Phys.~Rev. {\bf D69} 044006 (2004).

\bibitem{KL89} H.O.~Kreiss and J.~Lorentz, {\em Initial-Boundary
    problems and the Navier-Stokes equations}, Academic Press, New York
    (1989).

\bibitem{MoL} O.~A.~Liskovets, Differential equations I 1308-1323 (1965).

\end{thebibliography}

\end{document}